\documentclass[a4paper,rmp,onecolumn,11pt]{revtex4}

\usepackage[pdftex]{graphicx}
\usepackage{amssymb,amsfonts,amsmath}
\usepackage{epsfig}
\usepackage{natbib}

\long\def\symbolfootnote[#1]#2{\begingroup%
\def\thefootnote{\fnsymbol{footnote}}\footnote[#1]{#2}\endgroup}

\newtheorem{theo}{Theorem}

\newtheorem{prop}[theo]{Proposition}
\newtheorem{coroll}[theo]{Corollary}

\def\be{\begin{equation}}
\def\ee{\end{equation}}
\def\bc{\begin{center}}
\def\ec{\end{center}}
\def\bea{\begin{eqnarray}}
\def\eea{\end{eqnarray}}

\begin{document}

\title{Collaboration in Social Networks}
\author{Luca Dall'Asta$^{1,2,3}$, Matteo Marsili$^2$, and Paolo Pin$^4$}
\affiliation{$^1$Dipartimento di Fisica and Centre for Computational Sciences, Politecnico di Torino, Corso Duca degli Abruzzi 24, 10129 Torino, Italy\\
$^2$ The Abdus Salam International Center for Theoretical Physics, Strada Costiera 11, 34014 Trieste, Italy\\
$^3$ Collegio Carlo Alberto,  Via Real Collegio 30, 10024 Moncalieri (Torino), Italy\\
$^4$ Dipartimento di Economia Politica, Universit\'a degli Studi di Siena, Piazza San Francesco 7, 53100 Siena, Italy}

\begin{abstract}
The very notion of social network implies that linked individuals interact repeatedly with each other. This allows them not only to learn successful strategies and adapt to them, but also to condition their own behavior on the behavior of others, in a strategic forward looking manner. Game theory of repeated games shows that these circumstances are conducive to the emergence of collaboration in simple games of two players. We investigate the extension of this concept to the case where players are engaged in a local contribution game and show that rationality and credibility of threats identify a class of Nash equilibria -- that we call ``collaborative equilibria'' -- that have a precise interpretation in terms of sub-graphs of the social network. For large network games, the number of such equilibria is exponentially large in the number of players. When incentives to defect are small, equilibria are supported by local structures whereas when incentives exceed a threshold they acquire a non-local nature, which requires a ``critical mass'' of more than a given fraction of the players to collaborate. Therefore,  when incentives are high, an individual deviation typically causes the collapse of collaboration across the whole system. At the same time, higher incentives to defect typically support equilibria with a higher density of collaborators. The resulting picture conforms with several results in sociology and in the experimental literature on game theory, such as the prevalence of collaboration in denser groups and in the structural hubs of sparse networks.
\end{abstract}

\maketitle

The social network influences and constrains in non-trivial ways the behavior of individuals \citep{Jackson08} but also contributes to aspects generically referred to as {\em social capital} \citep{Homans58,Axelrod84,Sobel02}, which favor the emergence of coordinated actions or collaboration\footnote{We will explicitly refer to this endogenous profitable mutual exchange as {\em collaboration}, and not as {\em cooperation}, to stress the fact that it emerges from individual opportunistic incentives and not from group--based profit maximization, as in the cooperative games already introduced in \citep{VM44}.}.

The theoretical investigation about the emergence of collaboration and the fate of repeated actions on networks has, up to now,
mostly focused on adaptive learning, imitation and evolutionary game theory. The corresponding theoretical models \citep{Ellison94,ESS98,NM92,OHLN06,SPL06} seem to suggest that sustaining collaboration is easier on sparse networks and spatial structures than on dense groups.
On the contrary, there is no clear experimental evidence of such an effect. Results of recent experiments on voluntary contribution suggest instead that the establishment and maintenance of collaboration is easier on well connected groups than on sparse graphs \citep{KN07,C07,FMS10}, in some ways confirming the intuition behind the classical Coleman's closure argument \citep{C88}.
The picture becomes more complex as we move to analyze individual positions in the network: Experiments performed on star-shaped groups suggest that central nodes are fostered to collaborate more than peripherals \citep{FMS10}, corroborating another well-known idea,  i.e. Burt's argument on the importance of structural holes  \citep{B92}.

However, the network is not only the channel for the achievement of payoffs or the exchange of information, which are the key ingredients of learning and imitative behavior. The network also supports the establishment of trust, norms, contracts and other continuative collaborative relations in which the temporal dimension is crucial. In fact, a key aspect of a social network is that individuals connected by a link interact repeatedly with each other, calling into play forward looking strategic behavior typical of the theory of repeated games. This provides a formal framework for concepts such as threats, punishment and credibility, which have been so enlightening on the emergence of collaboration in simple setups (e.g. the prisoner's dilemma)  \citep{FM86}.

Here we investigate the extension of the theory of repeated games to network games. We focus on simple local contribution games which provide a perfect framework to highlight the main conceptual issues, and are closely linked to experimental works on voluntary contribution \citep{KN07,C07,FMS10,FG02}.
We start from the assumption, largely verified experimentally (see for instance Ref.~\citep{FG02}), that 
collaboration can be sustained only through a reciprocal relation of control and punishment, which plays an essential role in the theory of two-person repeated games \citep{FM86}. It follows that collaboration becomes conditional, i.e. an agent collaborates only if her peers also collaborate. We show that, in a multi-player setting, rationality imposes conditional collaboration to be {\em i)} reciprocal and {\em ii)} player-specific: Controlling only a subset of neighbors is enough to guarantee collaboration, and credibility of threats dictates that punishment should be limited to the minimal subset of neighbors which supports collaboration as a best response. These requirements identify a refinement of Nash equilibria, that we call Collaborative Equilibria, with a precise graph-theoretical interpretation. This turns the problem of finding and characterizing Nash equilibria into that of finding specific sub-graphs of the social network, which can be addressed by graph theoretical methods \citep{MPZ02,MM09}, and it provides considerable insight on the nature of possible collaborative structures which can be supported, depending on incentives and on the topology of the social network. In particular, we find that when the contribution cost is low, collaboration can be sustained by local commitments (corresponding to dimers or loops on the network). But when costs are high, collaboration requires non-local structures which span a finite fraction of the system. This has clear implications in terms of systemic fragility, because when costs are high a single individual defection may bring to the collapse of the whole collaborative network.

Our analysis of Collaborative Equilibria on ensembles of random graphs unveils a generic picture which, as will be discussed in the closing section, has many correspondences with well known results in sociology and in experimental economics.

\section{A Repeated Game of Local Contribution}
\label{sec-RGLC}
We focus on a class of local contribution games, recently popularized in Refs.~\citep{FG02,FC10}, which have the peculiarity that only the neighbors of a contributor and not the contributor herself benefit of the positive effects of a contribution. In this regard, we can interpret these local contribution games as a simple extension of the celebrated prisoners' dilemma to the case where there are more than two players, where the range of possible interactions between individuals is limited only to local ones by a fixed network structure, and where agents play the same strategy against each neighbor.

\subsection{The Model}
\label{sec-model}
Players occupy the nodes $N$ of an undirected social network and they interact with their neighbors in a local contribution game. Specifically, each player $i\in N$ has the option of either contributing ($s_i=1$) or not ($s_i=0$) to a local public good that has effect only on the neighbors of $i$ in $N_i$.\footnote{%
We adopt the notation $(N,\vec{N})$ to specify the social network, where $N$ is the set of nodes and the $i^{\rm th}$ element of $\vec{N}$, denoted $N_i$, is the subset of the neighbors of $i$ ($i\not\in N_i$. The social network is undirected: $i\in N_j$ implies $j\in N_i$).}
Contributing is costly, which means that players who contribute incur a cost $X_i>0$. At the same time, each player receives a unitary payoff from all agents $j\in N_i$ who contribute in her neighborhood. The payoff function of agent $i$ is then given by
\begin{equation}
\label{payoff}
\pi_i(s_i,s_{-i})=-X_i s_i+\sum_{j\in N_i} s_j
\end{equation}
where $s_{-i}$ stands for the vector of choices of all other players, except $i$.
It is clear that, in the single stage game, a positive $s_i$ is only a cost for player $i$, and hence defection (i.e. $s_i=0$) is a dominant strategy  for all $X_i>0$: $\pi_i(0,s_{-i})>\pi_i(1,s_{-i})$. If the game is played just once, the unique Nash equilibrium is one where $s_i=0$ for all $i\in N$.
When players are engaged in repeated plays of the game, with the same opponents, a much richer set of outcomes is possible. Indeed,
a continuum set of outcomes can be sustained as Nash equilibria \citep{FM86}.

Let the game above be played repeatedly at discrete times $t=0,1,2,3,\ldots$ and let $s_i^{(t)}$ be the option taken by agent $i$ at time $t$. A strategy now becomes an {\em action plan} $\sigma_i=\{s_i^{(t)},~t=0,1,2,\ldots\}$, which specifies the behavior of agent $i$ at all times. In particular, $s_i^{(t)}$ may be any function of the opponents' behavior $s_{-i}^{(t')}$ in the previous stage games $t'<t$. The payoff function is generalized to an inter-temporal utility function
\begin{equation}
\label{payoff1}
u_i(\sigma_i,\sigma_{-i})=(1-\delta)\sum_{t\ge 0} \pi\left(s_i^{(t)},s_{-i}^{(t)}\right)\delta^{t}
\end{equation}
where $\delta\in (0,1]$ is the factor by which agents discount future payoffs with respect to present ones.
For $\delta\to 0$ the game reverts to the single stage game with the unique non-collaborative equilibrium. The limit $\delta\to 1$ of (temporally) very far-sighted players is instead much more interesting, because players can choose from a huge (a priori infinite) space of possible strategies.

\subsection{Trigger Strategies}\label{sec-trigger}
To attack this problem, we start considering the simpler and well-known situation of a two-player game, where enforceable payoffs can be attained as Nash equilibria by reducing the set of possible strategies to a particular subset, known as {\em trigger strategies} \citep{R86}.
Trigger strategies encode the idea of {\em punishment}:  agent $i$ collaborates {\em a priori}, but if her  opponent $j$ misbehaves (i.e. $s_{j}^{(t)}=0$ for some $t\ge 0$), she will punish her with defective behavior for the infinite future (i.e. $s_i^{(t')}=0$ for all $t'>t$).  If the opponent is ``threatened'' in this way, her best reply, if $\delta$ is large enough and $X_i<1$, is to collaborate, or equivalently to use the same trigger strategy. This in fact guarantees the collaboration payoff $1-X_i>0$ in each round, as opposed to the zero payoff she would get if both defected. A crucial issue is that {\em threats of punishment must be credible}: a rational opponent will not consider credible a trigger strategy that inflicts a payoff loss to the player herself.

How does this insight carries over to multi--player settings, where players interact with their neighbors on the network? It is instructive to consider the case of three players, without loss of generality call them $1$, $2$ and $3$, all connected together in a closed triangle \footnote{%
From now on, for the sake of simplicity and without loss of generality, we  focus on the limit $\delta\to 1$ where the utility is dominated by the asymptotic behavior of players. Each statement derived for $\delta\to 1$ will be true for a $\delta<1$, which is large enough.},  with $X_i<1$. Imagine that players $1$ and $2$ are both using trigger strategies, making their collaboration conditional on the collaboration of the others. How should player $3$ behave? If she takes the threat of other players seriously, then she will also collaborate. But should she consider their threat credible? No, because if player $3$ were to defect, and both $1$ and $2$ also turn to defection, they would get a payoff of zero, whereas if they were to continue collaborating they would get a payoff of $1-X_i>0$ in each stage game. Hence the threat of players $1$ and $2$ is not credible and player $3$ better defects, free--riding with a payoff of $2$. On the contrary, if we assume $X_i > 1 $ $\forall i$ then full collaboration is individually rational and can be sustained in the repeated game as a Nash equilibrium outcome by means of credible threats. This example points out that generalizing trigger strategies entails making the strategies conditional on the behavior of a subset of the players: for $X_i<1$, player $1$ will punish player $2$ but not player $3$, because only one collaborator is needed in the neighborhood to make collaboration the best choice. In other words, {\em control by punishment is player-specific}\footnote{This notion was already implicitly present in several multi-player generalization of the Prisoner's Dilemma introduced to study the free-riding problem in groups of individuals \citep{GH93}.}.

Also, the same arrangement of players shows that {\em punishment should be reciprocal}. Imagine a situation where player $1$ punishes $2$, $2$ punishes $3$ and $3$ punishes $1$. Should $3$ abide to this arrangement? No. Again if she decides to defect, she may argue that it is not credible that $2$ will punish her, because again $2$ would end up with a lower payoff (zero) than what she would get by continuing collaborating with $1$.

\subsection{Collaborative Equilibria}\label{sec-ce}
These considerations generalize naturally to a notion of Nash equilibrium sustained by trigger strategies for a general network, with the {\em three conditions that punishment must be (1) credible, (2) player-specific, and (3) reciprocal}.
Let $\sigma=0$ be the ``always defect'' strategy and $\tau(\Delta)$
be a trigger strategy, conditional on the behavior of agents $j\in \Delta \subseteq N$. An agent playing $\tau(\Delta)$ will collaborate as long as all agents in $\Delta$ also collaborate.
Let $C\subseteq N$ be the subset of agents who play trigger strategies, whereas agents $i\not\in C$ always defect ($\sigma=0$). Individuals are only allowed to control their neighbors, i.e. if $i\in C$ plays  $\tau(\Delta_i)$, then $\Delta_i\subseteq N_i$.
It is convenient to introduce also the set of  {\em punishers} of $i\in C$, which is $\Gamma_i=\{j:i\in\Delta_j\}\subseteq C\cap N_i$. Then let
$c_i=|C\cap N_i|$ be the number of collaborators in $i$'s neighborhood and $\gamma_i=|\Gamma_i|$ be the number of punishers of $i$.
We assume that agents are (spatially) nearsighted in that, in considering the effect of deviations, they only consider direct effects of punishment.
Effects due to loops in cascades of defections and punishments are neglected\footnote{The defection of an agent $i$ or of one of its neighbors, could generate a cascade of defections due to punishment, which might lead also agents $j\in N_i\backslash \Gamma_i$ to defect. Such indirect effects of  punishment only occur if there are loops in the social network. If the social network is a tree, deviation of a neighbor of $i$ cannot have effects on other neighbors of $i$, if not through a change of $i$'s behavior. Hence nearsightedness is equivalent to assuming lack of common knowledge beyond the immediate neighborhood: players know only their neighbors and how many players each of them is interacting with.}.
Under these assumptions, we define the following refinement of Nash equilibria in the repeated local contribution game,

\begin{prop}[Collaborative Equilibria]
\label{prop1}
The arrangement where, for all $i\in C$, $X_i\le \gamma_i<X_i+1$ and $\Gamma_i=\Delta_i$, is a Nash equilibrium of the repeated game of nearsighted agents, for a sufficiently large discount factor $\delta$.
\end{prop}

\noindent
{\em Proof:}
If player $i\in C$ deviates by defecting, she can expect to receive a payoff per period of $c_i-\gamma_i$. If this is less than the payoff $c_i-X_i$ she gets by collaborating, then deviation is unprofitable. This occurs if $\gamma_i\ge X_i$, which is the first inequality in the proposition. Now, let $i,j\in C$ be neighbors and let $j\not\in \Gamma_i$ defect. Then, if $i$ continues to collaborate, her payoff is $c_i-1-X_i$ whereas by defecting $i$ gets $c_i-1-\gamma_i$ (note that since $j\not\in\Gamma_i$ the value of $\gamma_i$ does not change). If $i$ was collaborating before, i.e. if $\gamma_i\ge X_i$, then $i$ should continue collaborating, which means that $i$ should not punish $j$, i.e. $j\not\in\Delta_i$ or $i\not\in\Gamma_j$. As a consequence $i\in\Gamma_j$ implies $j\in\Gamma_i$, or $\Gamma_i=\Delta_i$, because otherwise the argument above would produce a contradiction.
Finally, let $i\in \Gamma_j$. If $j$ defects, $i$ should punish her, i.e. $i$ should also stop collaborating. This means that $i$'s payoff to collaborate ($c_i-1-X_i$) should be less that the payoff to defect, which is $c_i-1-(\gamma_i-1)$, because $j\in\Gamma_i$. This implies $\gamma_i<X_i+1$ which is the second part of the inequality in the proposition. $\Box$

\smallskip

Note that the assumption on spatial nearsightedness is not relevant for the incentive to collaborate. In other words, if $\gamma_i>X_i$ then collaborating would still be the best option if some other neighbors of $i\not\in\Gamma_i$ would also stop collaborating, as a consequence of indirect cascades of punishments. On the contrary, a situation where players were able to anticipate indirect defections may entail non-reciprocal control\footnote{Consider for example the case where $j\not\in \Gamma_i$ defects, and as a result of this $k\in\Gamma_i$ also defects in order to punish $j$ or some other neighbor. If $i$ continues collaborating she gets $c_i-2-X_i$ whereas if she defects her payoff is $c_i-2-(\gamma_i-1)$. Not punishing is not rational because $c_i-2-X_i>c_i-1-\gamma_i$ cannot be satisfied if $X_i\ge \gamma_i$. Hence $i\in\Gamma_j$.}.
Here, the assumption that agents cannot compute indirect effects of punishment is crucial.

Collaborative equilibria are a particular subset of all the Nash equilibria which can be supported by trigger strategies\footnote{%
It is possible to show that these equilibria can be defined as the equilibria of similar games with the same payoffs \citep{BK93,KH01}. For this we refer to a companion paper \citep{DMP11}.}. Here we keep the sophistication of the game theoretic treatment to a minimal level and we focus, instead, on the aspects related to the structure of collaborative equilibria and their computational complexity. In this respect, we note that a direct consequence of Proposition \ref{prop1} is that it turns the characterization of Nash equilibria into a graph theoretical problem:

\begin{coroll}
Every Collaborative Equilibrium of the local contribution game identifies a subgraph $(C,\vec{\Gamma})$ of the social network $(N,\vec{N})$, with $C\subseteq N$, $\Gamma_i\subseteq N_i$ and $|\Gamma_i|$ being the smallest integer larger than $X_i$. Every subgraph $(C,\vec{\Gamma})$ with these properties supports a Nash equilibrium where players $i\in C$ collaborate $(s_i^{(t)}=1)$ and players not in $C$ defect $(s_i^{(t)}=0)$.
\end{coroll}

\section{The Complexity of Collaboration on Networks}
\label{sec-complexity}
\begin{figure}[t]
\begin{center}
\includegraphics[width=6cm]{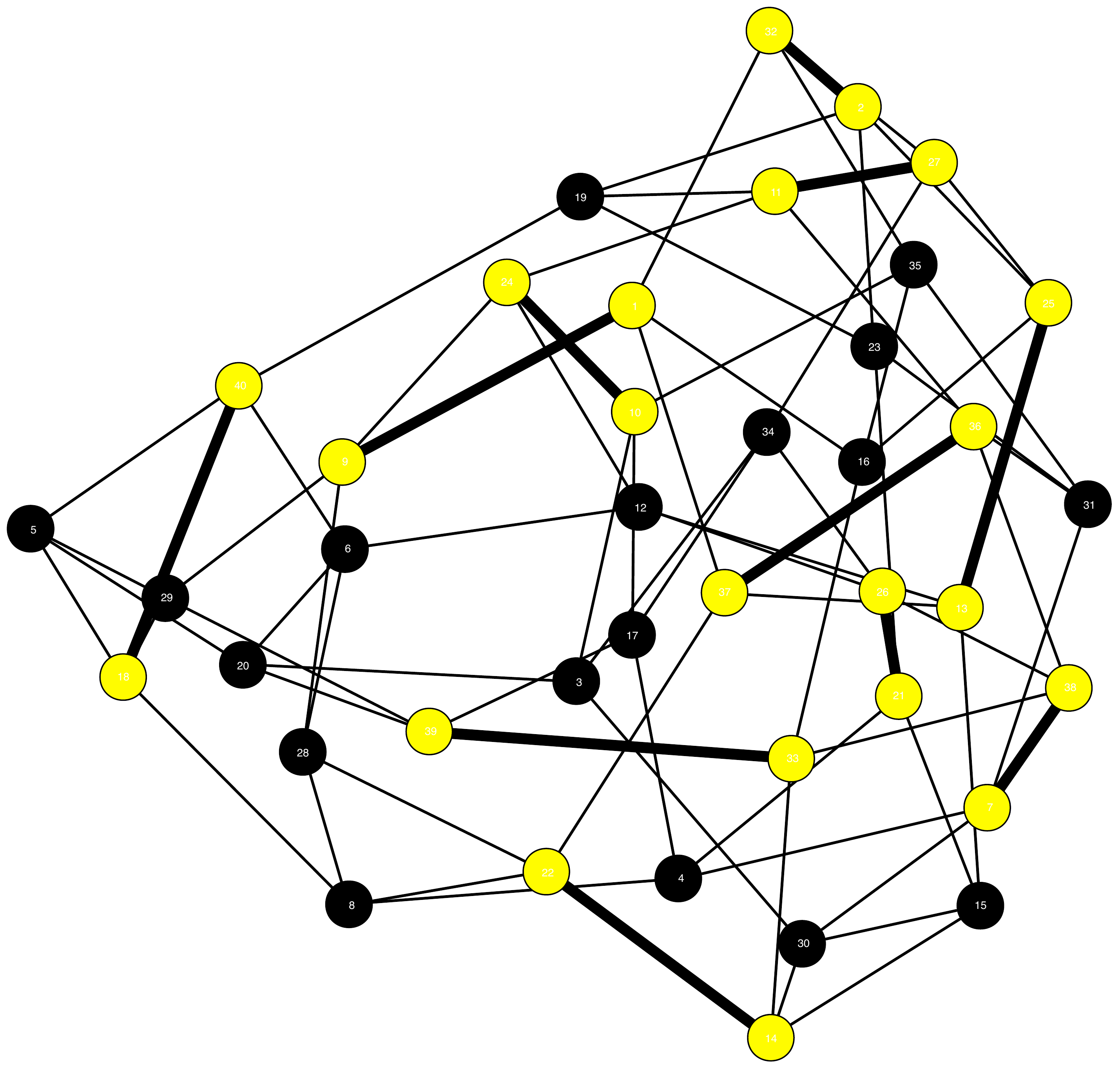}
\includegraphics[width=6cm]{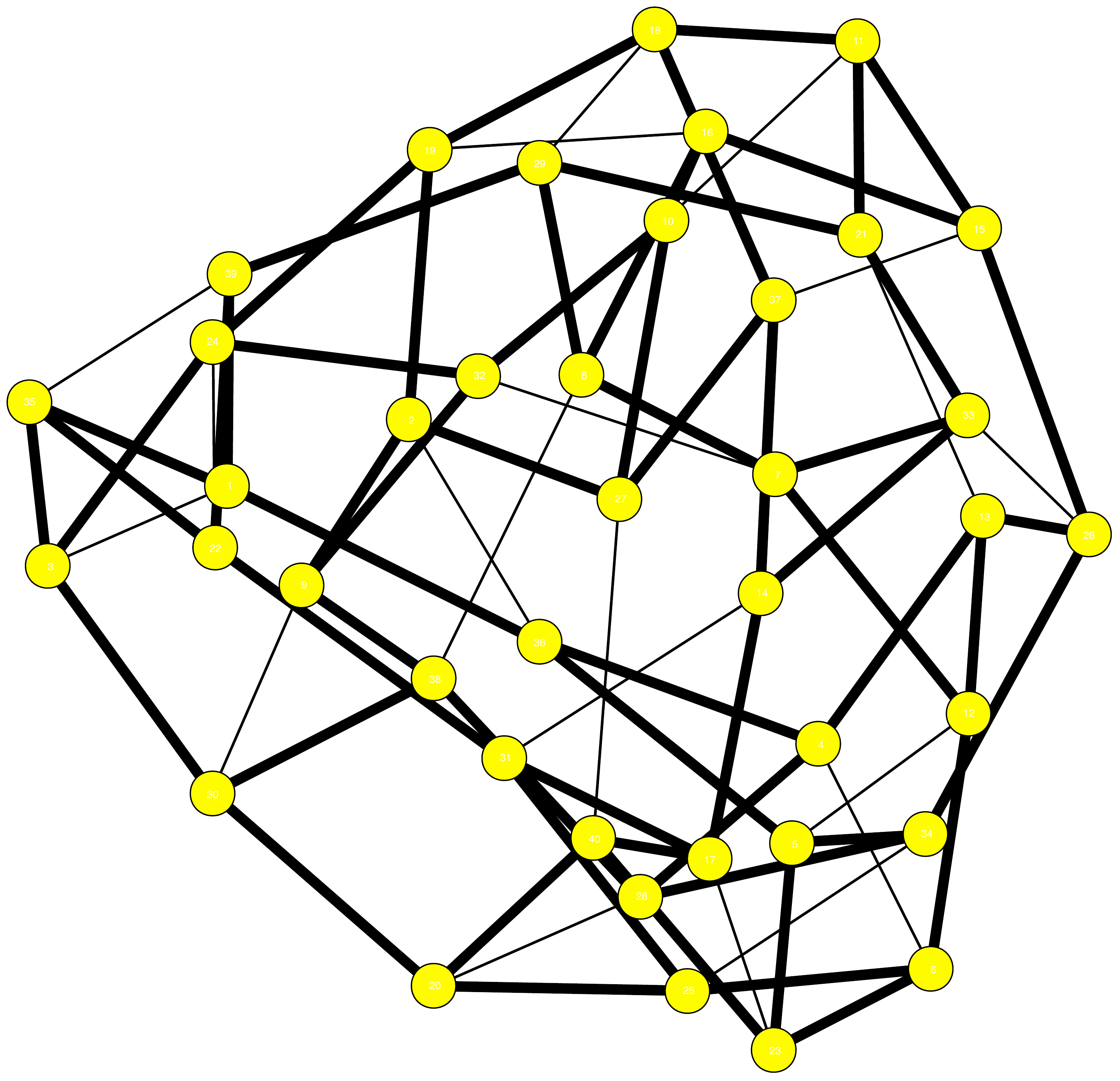}
\caption{Example of collaborative equilibria for $N=40$ players on a regular random graph ($|N_i|=K = 4$) with  $0<X_i<1, \forall i$ (left) and $2<X_i<3, \forall i$ (right) ). 
Defectors are in black whereas collaborators are in yellow. Thick links are those between players who collaborate conditionally on each other ($i\in\Gamma_j$ and $j\in\Gamma_i$).\label{fig1}}
\end{center}
\end{figure}

How many collaborative equilibria exist on a given social network and how does this number depends on the number of players, the payoffs and the network structure? How much hard is it to compute an equilibrium? How does the equilibrium ``respond'' to local perturbations?
The mapping of collaborative equilibria to collections of possibly disconnected subgraphs of a given socio-economic network allows us to address a number of questions. Some of these can be answered in very general terms: For example, if $0<X_i<1$ $\forall i$, the subgraphs which support collaborative equilibria are collections of disjoint ``dimers'' (see Fig.~\ref{fig1} left). The number of dimer covers is generically expected to increase exponentially with the size of the network, and local deviations have only local effects, as the defection of one player affects at most the behavior of the other player on the same dimer. If $1<X_i<2$ collaborative equilibria coincide with configurations of loops on the social network  and larger $X_i$ entail more complex structures (see Fig.~\ref{fig1} right). Correspondingly, the answer to the questions above becomes non-trivial. Also, the problem of finding subgraphs of a given network can be computationally very hard\footnote{For instance, in the case of random regular subgraphs, i.e.  $\gamma_i=\gamma$ $\forall i$, the problem  is known to be NP-complete on general graphs \citep{GJ79}.}, but it is typically very easy for dimers and loops, or for specific classes of graphs, such as planar graphs.

\subsection{Collaborative Equilibria on Random Graphs}
\label{sec-ac}
Quantitative predictions are possible for networks which are drawn from ensembles of random graphs. For large network sizes ($|N|\to \infty$) several properties attain a limit which is independent of the particular realization of the social network. We refer to these as {\em typical} properties, i.e. properties which are expected to hold with high probability for large networks. The local tree-like structure of large graphs with a finite degree makes the characterization of typical behavior tractable by means of message passing techniques \citep{MM09}, such as Belief Propagation (BP), which are exact on trees and are approximately correct also for finite, moderately clustered  graphs. The theoretical treatment is formally the same as in  \citep{MS06,PW06,MZ08}, which have considered similar graph-theoretic problems\footnote{All results presented here are obtained in the so-called replica-symmetric ansatz, that is under the assumption that the statistical properties of the equilibria are described by a unique Gibbs measure. This assumption is not always strictly correct, because the organization of the space of solutions could be more complex \citep{MPZ02}.
A signature that replica-symmetry could be a wrong assumption is provided by the stability of the BP fixed point. On single graph instances, this corresponds to the convergence of BP messages. In fact, we have observed that the BP equations are not always stable in the full range of values assumed by $\epsilon$.}.
BP allows one to derive results on ensembles of graphs in the limit $|N|\to\infty$  \citep{MM09} but it also provides efficient heuristic algorithms to find collaborative equilibria on given graph instances \citep{BMZ05,BZ06}. The messages which are exchanged in the BP algorithm are the probabilities $\mu_{i\to j}=P\{i\in\Gamma_j\bigcap C\}$ that player $i$ collaborates and punishes $j$, in the collaborative equilibrium\footnote{In the sub-graph problem, this is the probability that the link $(i,j)$ is part of the sub-graph, in the modified graph in which the link from $i$ to $j$ is removed. On trees, this operation disconnects the graph in subtrees.}. These probabilities, are updated through the equation \citep{PW06}
\begin{equation}
\label{BP}
\mu_{i\to j}=\frac{e^{-\epsilon} Z_{N_i\backslash j\to i}^{\gamma_i-1}}{Z_{N_i\backslash j\to i}^{0}+e^{-\epsilon} Z_{N_i\backslash j\to i}^{\gamma_i-1}+e^{-\epsilon} Z_{N_i\backslash j\to i}^{\gamma_i}}
\end{equation}
where, for any integer $q$ and subset $V\subseteq N$,
\begin{equation}
\label{BP2}
Z_{V\to i}^{q}=\sum_{U\subseteq V} \mathbb{I}_{|U|=q} \prod_{j\in U}\mu_{j\to i}\prod_{k\in V/U}(1-\mu_{k\to i})
\end{equation}
and the indicator function $\mathbb{I}_{|U|=q}$ restricts the sum only to subsets of $q$ elements.
In words, the numerator in the r.h.s. of Eq. (\ref{BP}), which is based on Eq. (\ref{BP2}), asserts that $i$ should control $j$ if there are $\gamma_i-1$ other players $k\neq j$ who control $i$. The denominator expresses the fact that, in a collaborative equilibrium, three situations are possible: 1) $i\not\in C$, and hence $i$ needs not be  controlled by any neighbor, 2) $i\in \Gamma_j$ and $i\in C$, or 3)  $i\not\in \Gamma_j$ and $i\in C$, i.e. $i$ is already controlled by $\gamma_i$ neighbors and does not need to control $j$. The parameter $\epsilon \in (-\infty,+\infty)$ is a statistical weight for collaborators, which is introduced to bias the distribution over equilibria towards those with a higher ($\epsilon<0$) or lower ($\epsilon>0$) density of collaborators.
The probability $P\{i\in C\}$ that a player $i$ collaborates or the probability $P\{i\in\Gamma_j\}$ that players $i,j\in C$ conditionally collaborate can be expressed in terms of the solution $\{\mu_{i\to j}\}$ of the set of equations (\ref{BP}-\ref{BP2}) as follows
\begin{eqnarray}
P\{i\in C\} & = & \frac{e^{-\epsilon} Z_{N_i\to i}^{\gamma_i}}{Z_{N_i\to i}^{0}+e^{-\epsilon} Z_{N_i\to i}^{\gamma_i}} \\
P\{i\in\Gamma_j\} & = & \frac{\mu_{i\to j} \mu_{j\to i}}{\mu_{i\to j}\mu_{j\to i}+(1-\mu_{i\to j})(1-\mu_{j\to i})}.
\end{eqnarray}

In practice, Eqs. (\ref{BP}-\ref{BP2}) can be iterated on a specific graph, substituting the value $\mu_{i\to j}$ with the value of the function on the r.h.s., until a fixed point is reached, which is guaranteed to occur on tree-like structures \cite{MM09}. The fixed point, however, is not unique. Indeed, for any collaborative equilibrium $(C,\vec\Gamma)$, binary messages $\mu_{i\to j}=\mathbb{I}_{i\in\Gamma_j\bigcap C}$ $\forall i,j$ are a solution of the BP equations. These ``pure strategy'' fixed points coexist with an internal solution, akin to a "mixed strategy" fixed point. As in similar problems \citep{MZ08}, the BP iteration converges to the internal solution, and makes possible a series of estimates on the statistical properties of the problem. In particular, we can compute the number $\mathcal{N}(\rho)$ of collaborative equilibria as a function of the density of collaborators $\rho=|C|/|N| \simeq \sum_{i \in N}P\{i\in C\}/|N|$. For large population size $|N|$, the number of equilibria is, to leading order, exponentially large in $|N|$, and one can define the entropy function $s(\rho)=\lim_{|N|\to\infty}\frac{1}{|N|}\log \mathcal{N}(\rho)$. Similarly, one can study the entropy of collaborative equilibria as a function of other parameters of the problem (e.g. costs of contribution) or topological property (e.g. average degree).
Moreover, simple adaptations of the message-passing algorithm, e.g. by iteratively fixing some variables (BP-decimation \citep{BMZ05}) or by introducing self-consistent biases on the messages (BP-reinforcement \citep{BZ06}), can be used to converge towards specific pure-strategy equilibria. In this respect, varying $\epsilon$ allows for searching collaborative equilibria with a given average density of collaborators $\rho$.

\begin{figure}[t]
\begin{center}
\includegraphics[width=10cm]{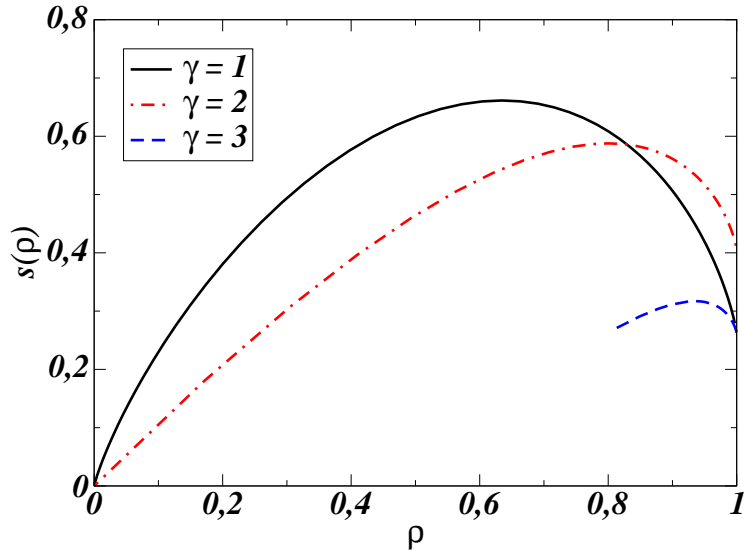}
\caption{Entropy $s(\rho)$ as a function of density $\rho$ of collaborators for $\gamma = \lceil X\rceil =1,2$ and $3$ on regular random graph with $|N_i|=K = 4$. Results are derived analytically for $|N|\to\infty$ as e.g. in Ref.~\protect\citep{MZ08}.\label{fig2}}
\end{center}
\end{figure} 

\subsection{Results}
\label{sec-res}
A clear picture of the properties of collaborative equilibria can be obtained considering the conceptually simple situation of regular subgraphs of random regular graphs. It corresponds to assume a uniform cost, i.e. $X_i = X$ $\forall i$, hence $\gamma_i=\gamma = \lceil X\rceil$, $\forall i$.
Examples of collaborative equilibria obtained by means of the BP-decimation algorithm on a random regular graph of degree $|N_i|=K=4$ are reported in Fig. \ref{fig1}. This suggests that the density of collaborators $\rho$ increases with the cost $X_i$. This conclusion is also supported by the behavior of the entropy $s(\rho)$, shown in Fig.~\ref{fig2}. This attains a maximum at a value $\rho_{typ}$ of the density of collaborators, which implies that, for large $|N|$,
almost all equilibria have $\rho\simeq \rho_{typ}$. Fig.~\ref{fig2} shows that $\rho_{typ}$ increases with $X_i$, suggesting that higher costs enforce higher densities of collaborators.
This is an apparently counterintuitive result that has however a simple explanation in the context of the model.
An agent collaborates only as long as there are enough neighbors around her that also collaborate, and that would punish her defecting if she defects: the higher the cost of collaboration, the more profitable it would be to defect, but hence also the more collaborating neighbors are needed to balance this incentive.

A remarkable feature of Fig.~\ref{fig2} is that while for $\gamma=1$ and $2$ equilibria exist for all densities $\rho$ of collaborators (i.e. $s(\rho)>0, \forall\rho\in(0,1]$), for $\gamma=3$ equilibria only exist for $\rho\ge \rho_c\approx 0.8$.
This reflects the fact that for $\gamma=1$, collaborative equilibria are collections of ``dimers" of collaborators who reciprocally control each others in pairs. For $\gamma=2$, equilibria are collections of loops, whose typical size  in a sparse random graph is $\mathcal{O}(\log |N|)$. For large graphs, collaborative equilibria with $\gamma_i=1$ or $2$ can be constructed adding and removing dimers or loops, adjusting the density $\rho$ in a continuous manner in $(0,1]$. For $\gamma=3$, instead, collaboration requires the formation of a regular subgraph of degree $3$, which is only possible if more than a fraction $\rho_c$ of the nodes are involved. Hence collaborative equilibria do not exist for $\rho<\rho_c$, and when they emerge (for $\rho>\rho_c$), they are exponentially many and they span a large fraction of the network. This has clear implications for the systemic stability of collaborative equilibria: while a local deviation for $\gamma<3$ only entails local rearrangements, for $\gamma\ge 3$ it is likely to cause the collapse of the whole collaborative structure.
Hence collaborative equilibria acquire a non-local character whereby the fate of agreements in a neighborhood depends on what happens in distant regions of the social network.
Such systemic fragility has its roots in the organization of the space of equilibria. Collaborative equilibria are formally obtained as solutions of constraint satisfaction problems (CSP) that belong to the class of locked CSP, recently introduced in Ref.\citep{MZ08}. In these CSP, if one modifies one variable in a solution, the rearrangement required to find another solution propagates across the network possibly affecting a sizable part of it. Here, for $\gamma < 3$, the shift from a given equilibrium to another one involves the rearrangement of at most $\mathcal{O}(\log |N|)$ nodes, i.e. a negligible fraction of the system. On the contrary, for $\gamma \geq 3$ the minimal distance between two equilibria is proportional to the number $|N|$ of nodes. 

\begin{figure}[t]
\begin{center}
\includegraphics[width=6cm]{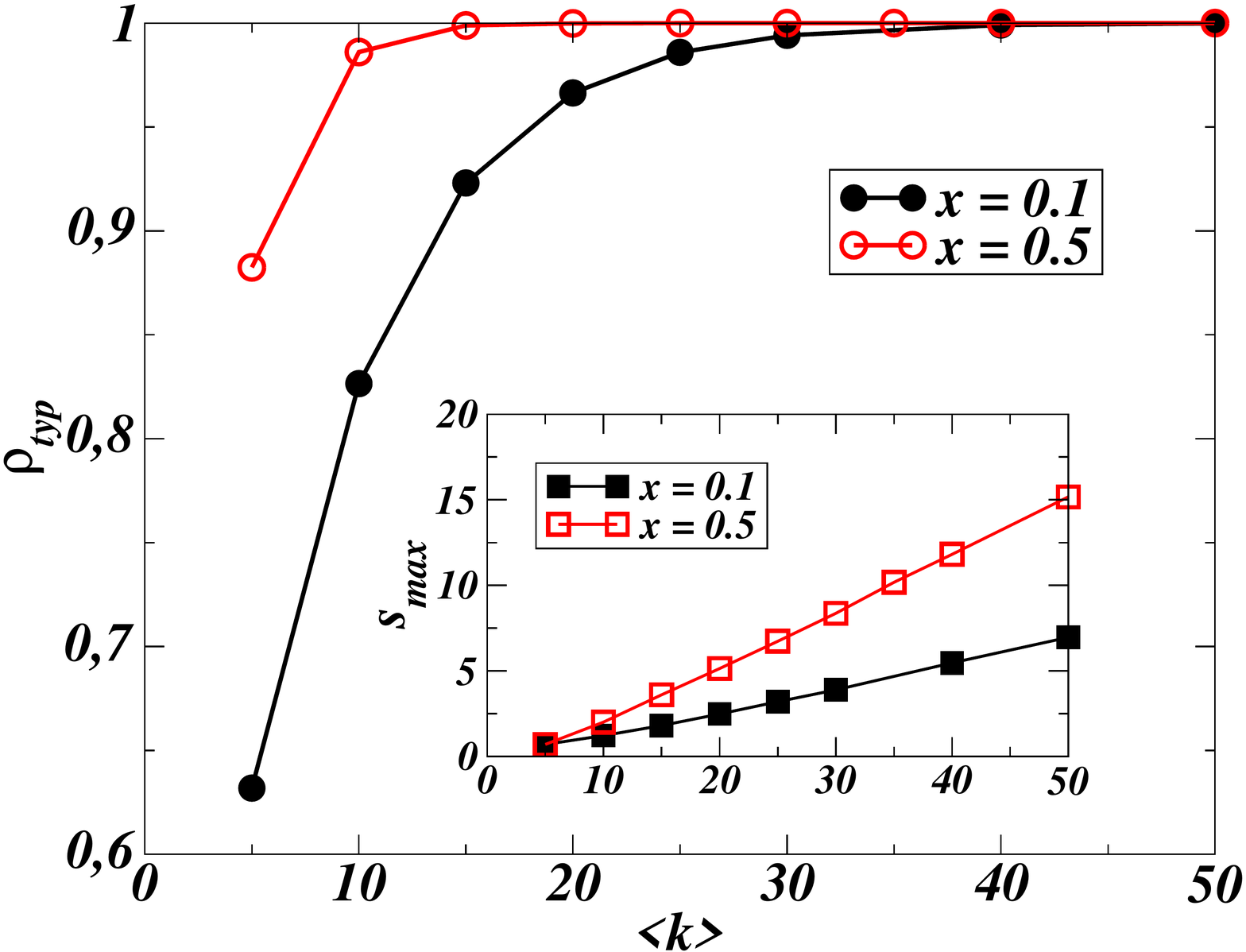}
\includegraphics[width=6cm]{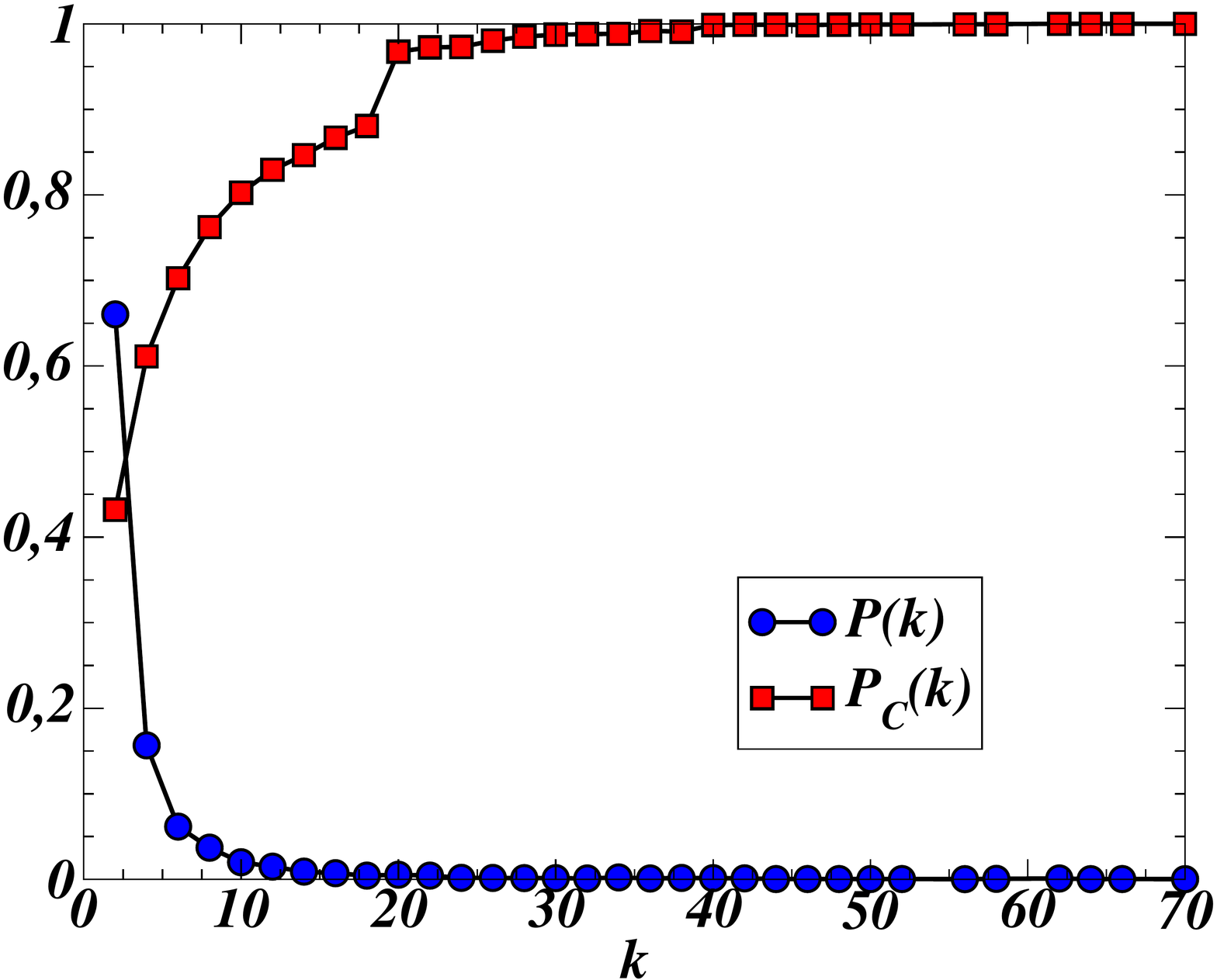}
\caption{(Left) Density $\rho_{typ}$ of collaborators in a typical equilibrium in Erd\"os-R\'enyi graphs as function of the average degree $\langle k \rangle$ for $x=0.1$ (black circles) and $0.5$ (red squares). The inset shows the corresponding typical entropies $s_{max}=s(\rho_{typ})$.
(Right) Probability $P_C(k)$ (red squares) that a node of degree $|N_i|=k$ collaborates in a typical collaborative equilibrium ($\rho_{typ}$) and low marginal cost ($x=0.05$) on an uncorrelated scale-free random graph of size $|N| = 5\cdot10^3$ with degree distribution $P(k) \propto k^{-2.5}$ for $2\le k\le 70$ (blue circles).\label{fig3}}
\end{center}
\end{figure} 

All the instances which have been analyzed confirmed the following set of generic features: {\em i)} on average the typical fraction $\rho_{typ}$ of collaborators increases with costs;  {\em ii)} the absence of equilibria at small densities $\rho$ for large costs; and {\em iii)} their fragility w.r.t. small perturbations and non-local character, for sufficiently large costs. These include Erd\"os-R\'enyi and scale free random graphs, with fixed costs $X_i=X$ (see also \citep{PW06}) and with costs of collaboration $X_i=x|N_i|$ that are proportional to the number of neighbors each player interacts with. In the latter case, when the marginal cost $x$ is small, equilibria are mainly formed by dimers and loops and can be found for any density $\rho$. When $x$ is large, non-trivial collaborative equilibria only exist for sufficiently large density of collaborators $\rho$, reproducing the ``critical mass" effect observed in regular random graphs. Interestingly, when the marginal cost exceeds a graph-dependent threshold $x_c$ ($x_c \simeq 0.79$ for Erd\"os-R\'enyi random graphs with average degree equal to 4), the number of equilibria vanishes in the full range of $\rho$ (i.e. $s(\rho)<0$), suggesting that the only possible equilibria are the all-defect or the fully collaborative (for $x=1$) ones.

In addition, we also found that {\em iv)} increasing the average degree promotes collaboration on average, because denser graphs admit typical equilibria of larger density $\rho_{typ}$ of collaborators, as shown in Fig.~\ref{fig3}(left) for Erd\"os-R\'enyi random graphs. The monotonic behavior of $s(\rho_{typ})$ with graph density (see the inset) is an evidence of the fact that in denser graphs there are much more possible ways of arranging a collaborating subgraph.
Finally, we found that {\em v)} within a collaborative equilibrium,
the more neighbors a player has, the more she is likely to collaborate (see Fig.~\ref{fig3} right).

\begin{figure}[t]
\begin{center}
\includegraphics[width=10cm,angle=0]{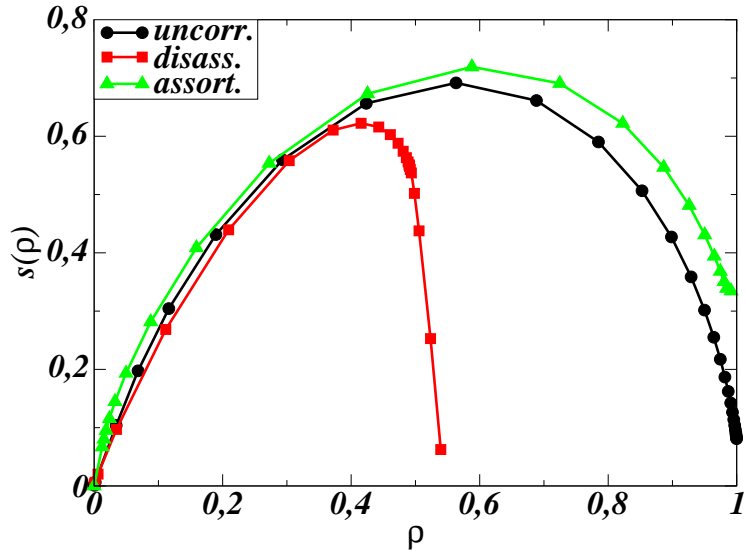}
\caption{Entropy $s(\rho)$ as a function of density $\rho$ of collaborators on correlated scale-free random graphs. We first considered an uncorrelated network ($J=0$ in Eq. 9 of Ref.\protect\citep{N07}) of $|N|=5\cdot10^3$ nodes, with degree distribution $P(k) \propto k^{-2.5}$ for $2\le k\le 70$ (black line and $\bullet$). Then we generated assortative/disassortative networks from this network, by means of a link-exchange Monte Carlo algorithm proposed in Ref.~\protect\citep{N07}, with positive ($J=0.1$, triangles) and negative ($J=-1$, squares) degree-correlation. The cost is $X_i=x|N_i|$ with  $x=0.1$.\label{fig4}}
\end{center}
\end{figure}

Social networks are far from being uncorrelated. Therefore it is important to consider the role played by degree correlations. To this end, we have generated assortative/disassortative networks starting from uncorrelated ones by means of a link-exchange Monte Carlo algorithm proposed in Ref.~\citep{N07}. The corresponding curves $s(\rho)$ computed for a moderate contribution cost $x=0.1$ are displayed in Fig.~\ref{fig4}. Our results suggest that {\em vi)} positive degree--correlation favors collaboration in that, the number of collaborative equilibria $s(\rho)$ and the typical fraction of collaborators $\rho_{typ}$ increases with degree correlation. Remarkably, in strongly disassortative networks, collaboration can be suppressed altogether for large $\rho$ (see Fig.~\ref{fig4}).


\section{Discussion}
\label{sec-discussion}

Experiments suggest that, apart from a very small fraction of innate altruists,  most individuals are self-interested, hence they rationally condition their own collaboration to that of their peers  \citep{FGF01}. However, in order to sustain {\em conditional collaboration} over time in a social group, credible punishment is necessary. In a network setting, we find out that agents condition their behavior only on those neighbors that are strictly necessary to get a higher payoff from collaborating than from defecting, i.e. punishment is player-specific. Immediately we find that control relationships between the agents have also to be reciprocal. Reciprocity is a familiar concept in the context of repeated games, especially in evolutionary game theory. In this respect, it is worth noting that our conditions to sustain collaboration completely agree with the concept of ``network reciprocity" introduced by Ohtsuki et al. \citep{OHLN06}, although resulting from a different assumption from  our forward--looking rationality.  Moreover, collaborative equilibria define a subgraph of pairwise interactions that reminds endogenous network formation games, in which stable coalitions depend on the existence of reciprocal pairwise interactions \citep{JW96,V06}.
Simply enforcing these three assumptions by means of trigger strategies on a repeated local contribution game on networks, we obtain our main conceptual result: {\em collaboration can be described in terms of a purely graph theoretical problem}.  We have called ``collaborative equilibria" the corresponding class of Nash equilibrium refinements.

This approach shows that the contribution cost (incentive to defect) has a major effect on the structure of the equilibria. When it is small, agents exert a low control on the neighbors and collaboration can be easily sustained in a repeated game. Increasing the cost/incentive, the system develops strong long-range correlations and {\em collaboration may require a critical mass}. Here, the absence of equilibria with low density of collaborators evokes Oliver and Marwell's ``critical mass theory", which is one of the most celebrated theories of collective actions \citep{MO93}. This maintains that simultaneous coordination of a sufficiently large subset of individuals is one of the possible solutions of the free-riding problem in situations in which individual contribution is extremely disadvantageous. Here we show that the existence of a critical mass emerges naturally from simple interaction, without the need of artificially introducing a threshold in the behavior of the agents. In this cases, collaboration is very fragile and, even if there are exponentially many equilibria, learning to coordinate on one of them can be very difficult.

An important conclusion from our analysis, that partially contradicts results from evolutionary game theory and learning dynamics based on imitation \citep{OHLN06,SPL06}, is that  network's sparseness and degree fluctuations do not necessarily favor the emergence of collaboration. In fact, the set of constraints imposed by control-and-punishment relations can be much more difficult to satisfy in sparse graphs than in dense groups. This is somehow in agreement  with experimental observation 
\citep{KN07,FMS10}, that measured systematically higher contribution levels in cliques and dense groups than in circular or linear arrangements.

The analysis of collaborative equilibria on different topologies suggests that the global properties of typical collaborative equilibria do not change considerably on different sparse networks. However, the local properties inside a network can be very different. This is evident in heterogeneous networks, where
{\em high-degree nodes have a considerably larger probability to collaborate than low degree ones}.
The result is in agreement with predictions of evolutionary game theory \citep{SPL06} and with experimental results obtained comparing the contribution levels of central and peripheral nodes in star-like graphs \citep{FMS10}. It also reminds the idea, firstly proposed by Haag and Lagunoff \citep{HL06}, of the existence of an uncollaborative fringe of agents connected to a collaborative core that can tolerate them.
This picture is particularly true when considering heterogeneous networks with assortative mixing.
On the contrary, in networks with negative degree correlations collaboration can be problematic, because of the mismatch between the conditions on high-degree nodes and the neighboring low-degree ones.

We conclude with a note on a recent experiment \citep{SW11} on the absence of social contagion with respect to  collaborative behavior. Social contagion refers to the idea \citep{FC10} that some personal behaviors (as the inclination to collaborate) may be transmitted via social networks. Suri and Watts \citep{SW11} find that increasing the number of collaborating neighbors does not directly imply a larger probability to collaborate (and vice versa).
Our model provides a theoretical foundation for such an observation. In fact, strategic conditional collaboration requires a precise number of collaborating neighbors and when the level of collaboration in the neighborhood is too high the temptation to free-ride takes over. On the other hand, for sufficiently large costs, a collaborative state can be destroyed because of a single deviation, that may induce a cascade effect over the network.

In summary, we have put forward a new framework to study the emergence of collaboration on networks, that is completely based on assumptions drawn from the experimental observations and on a rational forward-looking strategic behavior. Our results provide a theoretical explanation to a series of important empirical facts, and could motivate other experiments on repeated voluntary contribution games in networked systems. 
It provides insights in the analysis of strictly strategic problems, such as coalition formation, contractual agreements, and negotiation.




\begin{thebibliography}{99}
\bibitem{Jackson08}
Jackson M O (2008) {\em Social and Economic Networks} (Princeton, New York).

\bibitem{Homans58}
Homans C G (1958) Social Behavior as Exchange. {\em Am J Soc} 62: 597--606.

\bibitem{Axelrod84}
Axelrod R (1984) {\ The Evolution of collaboration} (Basic Books, New York).

\bibitem{Sobel02}
Sobel J (2002) Can We Trust Social Capital? {\em Journal of Economic Literature} 40: 139--154.

\bibitem{VM44}
von Neumann J, Morgenstern O (1944) {\em Theory of Games and Economic Behavior} (John Wiley and Sons, New York).

\bibitem{Ellison94}
Ellison G (1994) Collaboration in the Prisoner's Dilemma with Anonymous Random Matching. {\em Review of Economic Studies} 61: 567--588.

\bibitem{ESS98}
Eshel I, Samuelson L, Shaked A (1998) Altruists, Egoists, and Hooligans in a Local Interaction Model. {\em The American Economic Review} 88(1): 157--179.


\bibitem{NM92}
Nowak MA, May RM (1992) Evolutionary games and spatial chaos. Nature 359:
826Ð829.


\bibitem{OHLN06}
Ohtsuki H, Hauert C, Lieberman E, Nowak M A (2006) A simple rule for the evolution of collaboration on graphs. {\em Nature} 441: 502--505.

\bibitem{SPL06}
Santos F C, Pacheco J M, Lenaerts T (2006) Evolutionary dynamics of social
dilemmas in structured heterogeneous populations. {\em Proc Natl Acad Sci} 103: 3490--3494.


\bibitem{KN07}
Kirchkamp O, Nagel R (2007) Naive learning and collaboration in network experiments. {\em Games and Economic Behavior} 58 (2): 269--292.

\bibitem{C07}
Cassar A (2007) Coordination and collaboration in Local, Random and Small World Networks: Experimental Evidence. {\em Games and Economic Behavior} 58: 209Ð230.

\bibitem{FMS10}
Fatas E, Mel\'endez-Jim\'enez M A, Solaz H (2010) An experimental analysis of team production
in networks. {\em Exp Econ} 13: 399Ð411.

\bibitem{C88}
Coleman J S (1988) Social Capital in the Creation of Human Capital. {\em Amer. J. Sociology}
94: S95ÐS120.

\bibitem{B92}
Burt R S (1992) {\em Structural Holes: The Social
Structure of Competition} (Harvard U. Press, Cambridge).


\bibitem{FM86}
Fudenberg D, Maskin E (1986) The folk theorem in repeated games with discounting or
with incomplete information. {\em Econometrica} 54 (3):533--554..

\bibitem{FG02}
Fehr E, Gachter S (2002) Altruistic punishment in humans. {\em Nature} 415: 137--140.


\bibitem{MPZ02}
M\'ezard M, Parisi G, Zecchina R (2002) Analytic and Algorithmic Solution of Random Satisfiability Problems. {\em Science} 297: 812--815.

\bibitem{MM09} M\'ezard M, Montanari A (2009) {\em Information, Physics, and Computation} (Oxford University Press).

\bibitem{FC10}
Fowler J H, Christakis N A (2010) Collaborative behavior cascades in human social networks. {\em Proc Nat Acad Sci} 107(12): 5334--5338.

\bibitem{R86}
Rubinstein A (1986) Finite automata play the repeated prisoner's dilemma, Journal of Economic Theory, 39 (1), 83--96.


\bibitem{GH93}
Glance N S, Huberman B A (1993) The outbreak of collaboration. {\em Journal of Mathematical Sociology} 17(4): 281-302.

\bibitem{BK93}
Benoit J, Krishna V (1993) Renegotiation in Finitely Repeated Games. {\em Econometrica} 61 (2): 303--323.

\bibitem{KH01}
Kurzban R, Houser D (2001) Individual Differences in collaboration in Circular Public Goods Game. {\em European Journal  of Personality} 15: 37--52.


\bibitem{DMP11}
Dall'Asta L, Marsili M, Pin P (2011). Collaborative Equilibria in Networks, {\em Mimeo}.

\bibitem{GJ79} Garey M R, Johnson D S (1979). {\em Computers and Intractability} (Freeman, New York).

\bibitem{MS06} Marinari E, Semerjian G (2006) On the number of circuits in random graphs.
{\em J. Stat. Mech.} P06019.

\bibitem{PW06} Pretti M, Weigt M (2006) Sudden emergence of q-regular subgraphs in random graphs. {\em Europhys. Lett.} 75: 8.

\bibitem{MZ08} Zdeborov\'a L, M\'ezard M (2008) Constraint satisfaction problems with isolated solutions are hard. {\em J. Stat. Mech.} P12004.

\bibitem{BMZ05}
Braunstein A, M\'ezard M, Zecchina R (2005). Survey propagation: An algorithm for satisfiability. {\em Random Structures and Algorithms} 27: 201-226.

\bibitem{BZ06}
Braunstein A, Zecchina R (2006). Learning by Message Passing in Networks of Discrete Synapses. {\em Phys. Rev. Lett.} 96: 030201.

\bibitem{N07}
Noh J D (2007). Percolation transition in networks with degree-degree correlation. {\em Phys. Rev. E} 76: 026116.

\bibitem{FGF01}
Fischbacher U, G\"achter S, Fehr E (2001) Are people conditionally collaborative? Evidence from a public goods experiment. {\em Econ. Lett.} 71: 397--404.

\bibitem{JW96}
Jackson M O, Wolinsky A (1996) A Strategic Model of Social and Economic Networks. {\em Journal of Economic Theory} 71(1): 44--74.

\bibitem{V06}
Vega--Redondo F (2006) Building up social capital in a changing world, {\em Journal of Economic Dynamics and Control} 30: 2305--2338.

\bibitem{MO93}
Marwell G, Oliver P (1993). {\em The Critical Mass in Collective Action: A Micro--Social Theory}  (Cambridge University Press).

\bibitem{HL06}
Haag M, Lagunoff R (2006) Social norms, local interaction and neighborhood planning. {\em International Economic Review} 47(1): 265--296.

\bibitem{SW11}
Suri S, Watts D (2011) Collaboration and contagion in Networked Public Goods Experiments. {\em Mimeo}.


\end{thebibliography}
\end{document}